**RESEARCH ARTICLE**

# Can AI Rely on the Systematicity of Truth? The Challenge of Modelling Normative Domains

Matthieu Queloz[1]

**Abstract**
A key assumption fuelling optimism about the progress of Large Language Models (LLMs) in accurately and comprehensively modelling the world is that the truth is *systematic*: true statements about the world form a whole that is not just *consistent*, in that it contains no contradictions, but *coherent*, in that the truths are inferentially interlinked. This holds out the prospect that LLMs might in principle rely on that systematicity to fill in gaps and correct inaccuracies in the training data: consistency and coherence promise to facilitate progress towards *comprehensiveness* in an LLM's representation of the world. However, philosophers have identified compelling reasons to doubt that the truth is systematic across all domains of thought, arguing that in normative domains, in particular, the truth is largely asystematic. I argue that insofar as the truth in normative domains is asystematic, this renders it correspondingly harder for LLMs to make progress, because they cannot then leverage the systematicity of truth. And the less LLMs can rely on the systematicity of truth, the less we can rely on them to do our practical deliberation for us, because the very asystematicity of normative domains requires human agency to play a greater role in practical thought.

✉ Matthieu Queloz
matthieu.queloz@unibe.ch

1  Institute of Philosophy, University of Bern, Laenggassstrasse 49a, 3012 Bern, Switzerland

## 1 Introduction

Asked how Large Language Models (LLMs) could ever hope to comprehensively model the world as long as they were trained on incomplete and not fully accurate data, Dario Amodei, co-founder and CEO of Anthropic, invoked a time-hallowed idea as a reason for optimism, namely that *truths are integrated in a systematic web*:

> There's some relatively simple web of truth out there in the world … all the true things are connected in the world, whereas lies are kind of disconnected and don't fit into the web of everything else that's true. (Amodei, 2024)

We should not expect incompleteness and inaccuracies in the training data to be an insuperable obstacle to progress in AI, Amodei is suggesting, because LLMs could *in principle* work their way from those partially inconsistent data to the simpler, unified web of truth that underlies them, and rely on the systematicity of that web of truth to fill in lacunae and smooth out inaccuracies in the training data.

The crucial philosophical assumption that Amodei's optimism about overcoming gaps and inaccuracies in the training data depends on, however, is that the truth *is indeed systematic*: the totality of true statements about the world forms a whole that is not just *consistent*, in that it contains no contradictions, but *coherent*, in that the truths it consists of stand to each other in relations of rational support, allowing a given truth within the system to be inferred from other truths in the system. This coherence offers grounds for optimism, because it holds out the prospect that truths that are missing or misrepresented in the training data might be derived from the truths that are already represented in it. In other words, the consistency and coherence of the truth together facilitate progress towards *comprehensiveness* in LLMs' modelling of the world. If all truths interlock to form a systematic web, they promise, at least in principle, to allow LLMs to jump beyond the data they are trained on while acting as a safety net preserving them from error.

In practice, current LLMs likely leverage the systematicity of truth only to a limited extent, and only indirectly, insofar as it helps them achieve other objectives (see Section 3). The philosophical question, however, is how much leverage the systematicity of truth can provide in principle. Are all domains of thought such that progress in AI could in principle rely on the systematicity of truth?

The question is rendered more acute by the fact that philosophers have advanced various reasons to *doubt* that the truth is systematic across all domains of thought. In particular, value pluralists have offered forceful arguments to the effect that *in normative domains*, the truth is *not* necessarily systematic: statements expressing values, or describing how the world *ought to be* rather than how it is, do not necessarily fit together into a consistent, coherent whole. And insofar as the truth in normative domains is not systematic, this renders it correspondingly harder for LLMs to make progress in these domains, because they cannot then rely on the consistency and coherence of the truth to work towards comprehensiveness.

In a nutshell, my argument will therefore be the following: Where truth is systematic, progress towards comprehensiveness is *in principle* facilitated by the fact that LLMs can rely on the systematicity of truth to interpolate—and perhaps even extrapolate—from incomplete training data. But where truth is asystematic, progress is likely to be hampered by the fact that LLMs cannot rely to the same extent on the systematicity of truth to move beyond their training data. Especially—though not exclusively—in normative domains, there are reasons to think that the truth is largely asystematic. Therefore, there are reasons to think that LLMs will find it significantly harder to comprehensively model truths in normative domains. The less LLMs can rely on the systematicity of truth, moreover, the less we can rely on them to do our practical deliberation for us, because the very asystematicity of normative domains requires human agency to play a greater role in practical thought.

## 2 The Systematicity of Truth

The idea that truths interlock to form a systematically integrated whole has a long history in philosophy. The Stoics already referred to the *systema mundi*, the system of the world, and modern philosophers from Leibniz through Wolff, Lambert, and Kant to Hegel and Whitehead have elaborated this idea in various ways.[1] But what exactly does it mean to say that truths form a systematic web? The basic idea is that truths are not disconnected fragments. They interlock to form a unified whole— what the original compiler of the *Oxford English Dictionary*, James Murray, called "the fabric of fact."

This means, first of all, that truths are *consistent* with each other: they are free of contradictions. This includes both direct contradictions (P ∧ ¬P) and indirect contradictions, where the implications of a statement turn out to contradict each other ((P→Q) ∧ (P→¬Q)).

But consistency is a relatively weak requirement. A collection of disparate and unrelated truths might be free of contradictions without being systematically integrated. Moreover, consistency is easily achievable through the addition of further premises or qualifications (think of how, in the Ptolemaic model of the universe, the addition of ever more epicycles served to patch up observed inconsistencies).

When truths are systematically integrated, they are not merely consistent, but also *coherent*: they stand in relations of rational support to one another. This inferential interconnectedness reflects an underlying interconnectedness in the fabric of facts: the obtaining of one fact implies the obtaining of certain other facts and excludes the obtaining of yet other facts. If New York is east of San Francisco, then this implies that San Francisco is west of New York, and it excludes that New York is west of San Francisco. Accordingly, truths about geography must be rationally coordinated with each other. Acknowledging one truth about geography provides reasons for recognizing other truths about geography.

---

[1] The richest historical overview of the ideal of systematicity in philosophy is Otto Ritschl's *System und systematische Methode in der Geschichte des wissenschaftlichen Sprachgebrauchs und der philosophischen Methodologie* (1906). It is complemented by Messer (1907), and Rescher (1979, 3–8; 2005, 19–38) offers a lucid Anglophone account of the resulting picture. Losano (1968) and Troje (1969) also examine the development of the demand for systematicity, but in the context of law and jurisprudence. Stein (1968) sketches a brief history of the concept of system, and there are also various historical contextualizations of the notions of system at work in the thought of individual philosophers: Rescher (1981) examines the concept of system in Leibniz's work; Vieillard-Baron (1975) traces the concept of system from Leibniz to Condillac; Kambartel (1969) reconstructs the notions of system and justification in Kant's work, as do Rescher (2000, 64–98), Kitcher (1986), Guyer (2003, 2005), Abela (2006), and Ypi (2021); on Hegel and systematicity, see the essays in Brooks and Stein (2017), especially Thompson (2017). Franks (2005) explores the role of the demand for systematicity in German idealism.

As Nicholas Rescher (2005) has argued, this inferential interconnectedness introduces a useful form of *redundancy* into one's understanding of the world: given a sufficient number of truths about something, any one truth could be excised from the list and still remain derivable from the rest. Rescher (2005, 5) offers a perspicuous illustration using a simple tic-tac-toe-like situation:

|   |   |   |
|---|---|---|
| $x$ |   |   |
|   |   |   |

The following truths can be formulated about this situation, Rescher notes:

(1) There is exactly one $x$ in the configuration.
(2) This $x$ is not in the first row.
(3) This $x$ is not in the third row.
(4) This $x$ is not in the second column.
(5) This $x$ is not in the third column.
(6) This $x$ is not on a diagonal.
(7) This $x$ is not at column-row position (3, 2).

Were one to excise one truth from this list—say, (5), "This $x$ is not in the third column"—it could still be recovered from what remained. Excising (5) would still leave one able to infer it from (1), (2), (3) and (7).

As long as the fabric of fact is logically unified in this manner, a representation of that fabric of fact will thus contain redundancies that render it more *robust in the face of information gaps*: any truth that is lost, or perhaps even missing to begin with, can nonetheless be recovered from other truths about the fabric of fact.

This coherence of systematically integrated truths is relevant to the prospects for training LLMs, because it supports the *expansion of understanding by inference*. In particular, it promises to greatly facilitate *filling in what is missing* from the training data. Whether current LLMs can already reason their way to truths that are not explicitly present in their training data is contested. But the point is that the systematicity of truth supports this possibility *in principle*. Thanks to the systematic integration of the fabric of fact, truths that are not represented in the training data can in principle be recovered by inference. The systematicity of truth facilitates self-completion.

Up to a point, the coherence of systematically integrated truths also promises to help LLMs in *correcting inaccuracies* in the training data. This is what Amodei is driving at when he remarks that "lies are kind of disconnected and don't fit into the web of everything else that's true." If truths about a given domain are systematically integrated, then a candidate truth's *integrability with* what is already taken to be true can act as a *criterion* for whether the candidate truth should be accepted.

Of course, lies—or, more broadly, untruths—may, though disconnected from the web of what is in fact true, nonetheless be connected to other untruths. As any good liar knows, the hardest part of lying well is to preserve the appearance of systematic integrity across a web of lies. For, if the fabric of fact is systematically integrated, then the acceptance of an untruth will require corresponding changes in other parts of one's web of beliefs, which will require further adjustments in turn, and so on, so that even one false belief is capable of overturning many true beliefs, propagating falsity through the web.

Any supposition contrary to fact thus threatens to pervasively vitiate one's understanding of the world. This has been called *Burley's principle* (Rescher, 2005, 4), after the medieval philosopher Walter Burley, who observed that whenever a false contingent proposition is posited, any false proposition that is compatible with it can be derived from it (Kretzmann & Stump, 1989, 391). Given the acknowledgement of any two non-equivalent truths P and Q, we can derive from them the truth of ¬(¬P ∧ Q), which is logically equivalent to P ∨ ¬Q. But if we now assume only one untruth, say, ¬P, then there is no stopping there, because this at once implies ¬Q. Hence, it would seem that the acceptance of any one untruth (¬P) has the consequence that any other arbitrary truth (Q) has to be abandoned.

Burley's principle also has implications for LLMs. It highlights that besides the question of whether LLMs can rely on the systematicity of truth in principle, there is also the question of whether they succeed in relying on it in practice: do LLMs in fact manage to cotton on to the fabric of fact? Burley's principle underscores the real danger of their weaving together a web of falsehoods as a result of extrapolating even from as little as a single untruth. This reminds us of how easy it is for the web of truths to become obscured by a tissue of lies.

However, the fact that the assumption of a single untruth has systematic ramifications that are bound eventually to come into conflict with facts one knows to obtain can also be turned into an advantage in rendering LLMs truthful: it can be used to determine *which* systematically integrated web one should accept. Consider a detective trying to reconstruct a crime based on the testimony of various witnesses. Even if several of the witnesses are complicit and collude to entangle the detective in a systematic web of lies, that web of lies is likely to end up contradicting something that the detective, after a thorough investigation, comes to regard as incontrovertibly true. Even if the detective initially takes the misleading testimony at face value, therefore, the demand for all the facts of the case to fit together into a unified whole leads the web of lies to unravel in the end, because the systematicity of truth gives the detective *critical leverage* over the epistemic authority of the witnesses:

it enables the detective to retroactively re-evaluate the trustworthiness of witnesses in light of how well their testimony *fits in* with what gradually emerges as the most consistent and coherent account of what happened.

Analogously, LLMs could *in principle* rely on the systematicity of truth to retroactively re-evaluate inaccuracies in their training data in light of their overall fit with the most consistent and coherent web of statements about the world. And when competing webs are equally consistent and coherent, components of the training data that are given special weight as certifiably authoritative sources can be treated as ground truths and act as tie-breakers. Thanks to the systematicity of truth, moreover, these high-quality inputs can be used not merely to overturn statements that directly contradict them, but can be leveraged more widely. When systematically integrated, their implications can reveal tensions with the implications even of seemingly unrelated statements. The systematicity of truth, especially when supported by such a reliable basis, thus facilitates self-correction.

The upshot is that the systematicity of truth promises to be a significant aid to the development of comprehensive LLMs, because that systematicity is something that LLMs can in principle rely on for *self-completion* and *self-correction*. This raises the question of how things currently are in practice: to what extent are LLMs already being trained to be sensitive to consistence and coherence?

## 3 The Role of Consistency and Coherence in Training LLMs

Present-day LLMs likely leverage the systematicity of truth only to a limited extent, and only indirectly, insofar as it helps them achieve training objectives aimed at something else. I say "likely" because there are two sources of uncertainty: one is the obvious one—leading AI firms tend to keep details of the training process under wraps for as long as it gives them a competitive advantage; but the other is more philosophically interesting: what these models are *trained* to do is one thing, and what they *actually* do in order to do what they are trained to do is quite another. Based on what is publicly known about the training process, one can determine to what extent LLMs are humanly *designed* to be sensitive to consistency and coherence; but one can only speculate, at this stage, about the precise extent to which they in fact latch on to wider patterns of consistency and coherence as a means of meeting the training objectives defined by software engineers.

The training process that LLMs normally undergo can be broken down into two phases: a *pre-training* phase and a *post-training* phase. The purpose of the pre-training phase is to turn the model into a linguistically competent repository of general knowledge (this need not imply that the model itself knows things; Wikipedia is a repository of general knowledge even if Wikipedia does not itself know things).

An LLM can be thought of as a mathematical function that takes a sequence of words as input and outputs a probability distribution over possible next words, from which a likely next word is then selected. More precisely, so-called "autoregressive" LMMs, such as OpenAI's GPT models, Google's Gemini models, or Anthropic's Claude models predict the next token in a text string (which could be a word or a smaller chunk of text, such as a punctuation mark) based only on preceding tokens—as many as the model's

"context window" (roughly, its working memory) will allow.[2] At the start of the pre-training phase, the model's parameters are set at random and the output is wildly off. But the model learns to predict likely continuations of a text string by training on colossal amounts of text scraped from the internet and internalising patterns in the co-occurrence of words: e.g. "the best type of pet is a" is followed by "dog" 59.49% of the time, by "cat" 18.74% of the time, and by "horse" only 0.72% of the time.[3]

This is done through "self-supervised learning" (SSL), meaning that humans do not structure and label the data to help the model interpret it. Instead, they set a training objective: to minimise the divergence between the probability distribution over possible next tokens and the actual next token. This objective is coded in through a mathematical function that penalises the model for assigning a low probability to the correct token. The amount of divergence provides the model with a feedback signal it can learn from using a "backpropagation" algorithm, which nudges each parameter in the direction that promises to improve future predictions. Over time, this calibrates the parameters to yield good next-token predictions. The parameters are tweaked until they form computational circuits whose execution replicates the probabilistic relationships observed in the training data. Aligning the model's predictive dispositions with these statistical patterns in natural language lays the groundwork for the model's ability to formulate grammatically and contextually plausible-sounding sentences. It also makes the model more likely to output true statements on a wide range of topics. In that sense, SSL can be said to embed a vast number of facts into the model.

The training objective in SSL—next-token prediction—appears, in itself, indifferent to consistency and coherence. But *what* LLMs are trained to do reveals little about *how* they learn to do it (Goldstein & Levinstein, 2024; Herrmann & Levinstein, 2025; Levinstein & Herrmann, 2024). And how they learn to do it is something that remains in significant respects opaque to us (Beisbart forthcoming-b, a). We design the training process and define the training objective, but the capabilities that emerge through training are more *grown* than designed; the training objective is like the light towards which the model grows, leaving it to the model to develop strategies it can leverage to grow in that direction.

Consequently, it is difficult to *rule out* that these models learn to rely on patterns of consistency or coherence to better achieve their training objective. Conceptualising them as doing *nothing but* next-token prediction—where that implies that they rely only on shallow statistical patterns like n-gram likelihoods—risks overlooking "emergent abilities," i.e. abilities that are not explicitly programmed in, but that the models develop on their own.[4] They might get better at next-token prediction *by* learning to pursue subordinate objectives that yield good proxies for next-token prediction. After all, considering whether a certain word would be a consistent and coherent way to continue a text string seems like a promising way to track its probability of being the actual continuation. However, the limited consistency and coherence of the outputs achievable through pre-training alone—there is a reason it is called "pre-training"—suggests that pre-training only instils a rudimentary form of sensitivity to consistency and coherence—one that remains highly *localised* to individual sentences or perhaps paragraphs, but that breaks down over longer ranges or across different conversational threads (Elazar et al., 2021; Jang & Lukasiewicz, 2023; Kumar & Joshi, 2022; Liu et al., 2024). Moreover, some inconsistencies and incoherences in the pre-training data (e.g., conflicting statements) get embedded into the model along with the facts.

---

[2] This so-called "causal language modelling" contrasts with "masked language modelling," where models such as BERT are trained to predict a masked word given both preceding and subsequent words.

[3] These are actual probabilities from GPT 3.5.

[4] On emergent abilities, see Wei et al. (2022). On why conceptualising these models as nothing but next-token predictors is overly reductive, see also Downes, Forber, and Grzankowski (2024).

When we interact with an LLM, we do not just want *probable* continuations of text strings, but *accurate*, *informative*, *concise*, and *polite* ones, to name but a few desiderata. A pre-trained model accordingly needs to be *fine-tuned* for specific tasks, such as instruction following, conversation, question-answering, summarisation, translation, or coding. The purpose of the post-training phase is to refine the somewhat rough, "pre-trained" model to imitate the patterns that characterise exemplary completions of these various tasks. This can be thought of as a form of imitation learning.

A standard technique for this is "supervised fine-tuning" (SFT), where the objective typically remains next-token prediction, but the training data is swapped for smaller, task-specific datasets of input–output pairs that have been carefully curated, structured, and labelled by humans. These could be ideal examples of customer service dialogues, for instance, or exemplary pairs of medical questions and answers. By training the model to do next-token prediction on those ideal examples, it is implicitly encouraged to emulate the *virtues* of these examples—not just tone, length, and style, but also characteristics like epistemic humility or nuance.

Again, consistency and coherence are not themselves training objectives in SFT. Yet the exemplary input–output pairs are presumably chosen, among many other reasons, *because* they are paradigmatic examples of consistency and coherence. By being fine-tuned to follow these templates, the model is therefore indirectly being trained to be more sensitive to consistency and coherence when it generates its own responses to user prompts.

However, the considerations of consistency and coherence that inform the assembly of task-specific datasets remain limited to consistency and coherence *within* a given input-output pair and to that input-output pair's consistency and coherence *with the truth*—or rather, what those who put together the datasets took to be the truth. And it is not clear that training to do next-token prediction on datasets that exhibit these features does much to train an LLM to leverage the consistency and coherence of the fabric of fact more widely. It may make the LLM more likely to produce internally consistent and coherent outputs, but there remains an important difference between producing an internally consistent and coherent output and relying on the consistency and coherence of the wider fabric of fact to generate that output. Output-consistency or output-coherence are not the same as consistency-based or coherence-based outputs.

To further align the model's behaviour with human expectations, preferences, and values that are hard to formalise, a common technique is "reinforcement learning from human feedback" (RLHF).[5] The basic idea is to have human annotators rate or rank the model's outputs according to which ones they prefer. The annotators might be instructed to select the answers they find more *helpful*, *honest*, and *harmless*, for instance, as Amanda Askell, a philosopher in charge of fine-tuning the character traits of models at Anthropic, put it in Askell et al. (2021), a paper that also informed OpenAI's RLHF efforts (Ouyang et al., 2022). This feedback can be used to reinforce the model's tendency to generate output that scores highly along these dimensions. To do this at scale, the human feedback is used to train a separate model—a "reward model"—to encode those human preferences. This reward model can then be used to score the outputs of the language model, reinforcing those that score highly.

Sensitivity to consistency and coherence is not the primary objective of RLHF either, since the LLM's training objective during RLHF is to maximise the score given to it by the reward model. But since human annotators are apt to penalise not just inconsistency or incoherence *within* an output or an input-output pair, but also inconsistency or incoherence with what they take to be the truth, RLHF indirectly incentivises sensitivity to consistency and coherence at least in these two significant respects.

The result of all this training are models that are increasingly good at ensuring consistency and coherence within a given response. But this hard-won achievement still tends to fracture as conversations get longer. Among "the most disconcerting weaknesses of these models' performance," Cameron Buckner observes, are their "tendency to meander incoherently in longer conversations and their inability to manifest a coherent individual perspective" (2023, 283). This experience of disintegration over long ranges erodes trust in the output. We find it harder to accept someone's assertion that p if they also assert in the same conversation that not-p. Indeed, the inability to manifest consistent and coherent propositional attitudes undermines the impression that we are dealing with propositional attitudes at all. "With regard to beliefs," Bernard Williams notes, "it is not simply that a person will seem inconsistent or contradictory or hopeless if they change too often; rather, if they change too often for internal reasons, they will not be beliefs but rather something like propositional moods" (2002, 191).

There is growing recognition that to make models more consistent and coherent over long ranges, one might need to fine-tune models *directly* for logical properties like consistency or coherence. Both fine-tuning for consistency within the model's output as well as fine-tuning for consistency with external knowledge databases are methods that are being explored (Liu et al., 2024; Zhao et al., 2024; Zhou et al. 2024). What remains largely untapped, however, is the potential of measuring and optimising directly for the far more general ability that Amodei is gesturing at: the ability not just to produce systematic outputs, but to leverage the wider systematicity of truth to overcome gaps and inaccuracies in the training data.

---

[5] Though newer techniques keep being developed, and AI labs increasingly use complex combinations of these and other fine-tuning methods. In "reinforcement learning from AI feedback" (RLAIF), for example, human labellers are replaced by existing AI models that act as "graders." In "constitutional AI" approaches, the model relies on a predefined set of principles (a "constitution") to critique its own outputs. See, e.g., Bai et al. (2022) and Lee et al. (2023).

Thus, while it is plausible that current training methods like SSL, SFT, or RLHF indirectly foster some sensitivity to consistency and coherence, that sensitivity appears limited in scope, and does not seem to leverage the wider systematicity of truth. Future research could explore methods that directly train LLMs to exploit the consistency and coherence of the fabric of fact for self-completion and self-correction—perhaps in conjunction with Retrieval-Augmented Generation (RAG), which seeks to integrate text generation with information retrieval from reliable sources. Such a broadened sensitivity to the systematic integration of the truth would in turn help models be more consistent and coherent across sessions over time. At present, LLMs are still too much like the press as described by Karl Kraus: indifferent to whether what they write today contradicts what they wrote yesterday.

Even if better training techniques can overcome these shortcomings, however, a more fundamental challenge awaits: truths in certain domains may themselves be asystematic.[6] As we shall see in the next section, various philosophers have pointed out that *normative* domains, especially, which notably encompass ethics and politics, form complex landscapes of often conflicting values, ideals, virtues, and principles. This has implications for the prospects of LLMs in these domains.

## 4 Value Pluralism and the Asystematicity of Normative Domains

The tradition of *value pluralism*, whose relevance to AI was recently underlined by John Tasioulas (2022), offers forceful arguments to the effect that when it comes to truths about what is valuable, the truth may well be asystematic: statements expressing values, or describing how the world *ought to be* rather than how it is, do not necessarily fit together into a consistent, coherent whole.

For our purposes, value pluralism is best understood as involving four claims. First, in contrast to monism, which holds that there is really only *one* thing that is intrinsically valuable (hence "monism"), pluralists hold that there is not just one, but a *plurality* of irreducibly distinct values. A paradigmatic form of value monism is utilitarianism, which takes the overarching master value in terms of which everything else is ultimately measured to be some form of utility, such as preference satisfaction.[7] The influential computer scientist Stuart Russell (2019), for example, advocates such a preference-based utilitarianism as the best-suited ethical framework for aligning AI models with human values.

---

[6] Cummins et al. have also made the congenial observation that while "some domains are cognized via a grasp of their underlying structure," "some domains are cognized without grasping any significant underlying structure" (2001, 174–175). However, their conception of domains and what it takes for them to be structured differs starkly from the one at issue in what follows.

[7] At least, this applies to the best-known elaborations of utilitarianism. As Sen (1981) argues, it is in principle possible to envisage a value pluralist elaboration of utilitarianism.

Second, pluralists maintain that there is, in many cases, an irreducible *incompatibility* between these different values. As Bernard Williams articulates the view, pluralists consider it a "deep error" to suppose "that all goods, all virtues, all ideals are compatible, and that what is desirable can ultimately be united into a harmonious whole without loss" (Williams, 2013, xxxv). Our values are bound to end up pulling in competing directions when pursued in concert, not merely because time is short or the world recalcitrant, but because the values themselves inherently conflict.[8] We face inevitable trade-offs between diverging ends, the realisation of some of which can only be obtained at the expense of others (Berlin, 2002, 213–214). This is not just the Rawlsian claim that the values of some groups in society clash with the values of other groups. It is the stronger claim that even the values of a single individual can present one with fundamentally irresoluble conflicts: think of the notorious tension between the ancient virtue of excellence and the Christian virtue of humility; or of the tensions between truthfulness and kindness, liberty and equality, loyalty and honesty, tradition and progress, justice and mercy, or security and privacy.

This commitment to irreducible incompatibility contrasts with views on which there are no truly irreducible value conflicts, or no truly ineluctable moral dilemmas: any apparently conflicting demands can ultimately be resolved by showing that one kind of demand is lexically prior to the other, or that one demand was merely apparent. One example is the Kantian view on which morality ultimately makes no conflicting claims and nonmoral claims can be discounted. This immunization against conflicting moral obligations turns on the assumption that *ought* implies *can*: one is only under a moral obligation to fulfil claims of morality that *can* be fulfilled together, and what remains unfulfilled cannot really have been a moral obligation to begin with, but must have been a merely *apparent* obligation, or at best a *prima facie* obligation. Utilitarianism, for its part, avoids irreducible incompatibility by insisting that there is really only *one* moral obligation, which is to maximise utility. It treats any apparent value conflict as resolvable by regarding the values involved as *versions* or *aspects* of one and the same value—utility—and incorporates them into the utility calculus to determine what is optimific overall. (Some indirect forms of utilitarianism try to accommodate genuine conflicts between distinct values by emphasising the autonomy of our first-order values; but insofar as second-order considerations of utility remain the only real validation for these first-order ways of thinking, that combination notoriously threatens to unravel under reflection.) It is this promise of compatibility or "consistency" between our values that draws psychologist and neuroscientist Joshua Greene towards a utilitarian approach to AI alignment: "Before we put our values into machines," he writes, "we have to figure out how to make our values clear and consistent" (2016, 1515), and utilitarianism promises to help with that by providing "a unified system for weighing values" (2013, 15).

---

[8] See Berlin (2013, 12), Berlin and Williams (1994), and Queloz (2025). Further elaborations of the pluralist outlook include Larmore (1987), Stocker (1990), Kekes (1993), Chang (1997a, 2015a), Dancy (2004), and Hämäläinen (2009), 548). See also Chang (2015b), Heathwood (2015), Mason (2023), and Blum (2023) for overviews.

This leads to the third pluralist claim, that different values are in many cases *incommensurable*. When values conflict, there is no *common currency* in terms of which to compute the gains and losses involved in trading one value against another. But it also means something wider, namely that "there is no other determinate and general procedure for solving conflicts, such as a lexical priority rule" (Berlin & Williams, 1994, 306). Resisting the pressure to come up with techniques for making incommensurable values commensurable, pluralists hold that our efforts "should rather be devoted to learning—or learning again, perhaps—how to think intelligently about conflicts of values which are incommensurable" (Williams, 2001, 89).

One important step in that direction is to realize that while commensurability implies comparability, *incommensurability* does not imply *incomparability*. "Comparison," Ruth Chang emphasizes, "does not require any single scale of units of value" (1997b).[9] Moreover, the lack of a single scale of value does not render arbitrary an agent's judgements concerning which of two incommensurable values is more important in a given connection. Judgements of importance need not be any less rational or reasonable simply because they do not rely on a common currency of value. One can still have *reasons* to think that one value should prevail over the other in a given situation—it is merely that these reasons will not take the form of a common currency or a lexical priority rule, and will be more context-sensitive than these highly general procedures for solving value conflicts.[10]

Fourth and finally, pluralists hold that we should not necessarily expect *truths about* values to interlock in a systematically integrated whole. If many of our different values are genuinely distinct, incompatible, and incommensurable, the relationship between the truths about these values becomes correspondingly complex and conflictual. As Thomas Nagel "observes, "truth in science, in mathematics, or in history has to fit together in a consistent system," but "our evaluative beliefs are not part of the attempt to describe a single world" (2001, 108–9). Floating free of the attempt to describe a single world, truths about values are at greater liberty to be radically inconsistent and incoherent.

Of course, it is not just in normative domains that the unity and systematicity of truth has been questioned. Philosophers of science like Nancy Cartwright (1983, 1999) have argued that even truths *in science* do not all fit together into a single, pyramid-shaped system with physics at the top. On her view, science resembles a patchwork more than a pyramid, forming a mosaic of models suited to particular domains rather than a grand unified theory.[11] Even Otto Neurath, the figurehead of the Vienna Circle's Unity of Science Movement, urged scientists to abandon their belief in "the" system of science into which all truths about natural phenomena can be fitted (Neurath, 1935, 17).

---

[9] Chang (2015b, 25) maintains that even monism does not strictly entail comparability, because different *qualities* of a single value need not be comparable; indeed, on her account, even different *quantities* of a single value need not be comparable.

[10] This line of thought has been elaborated notably by Dancy (1995, 2004) and the large literature on particularism; see the essays in Hooker and Little (2000) for a good overview.

[11] See also Feyerabend (1993), Dupré (1995), and the collection edited by Galison and Stump (1996). For a more recent and partly critical examination of studies of inconsistency in science, see Vickers (2013). For an examination of the relation of Cartwright's account to Dancy's particularism in ethics, see Sandis (2006).

But truths in science still exhibit a great deal of *localised* consistency and coherence: within a scientific discipline, or within a model, they largely fit together into a systematic whole.[12] Moreover, though they latch onto different properties at different levels of abstraction, they are still understood to be different models *of one natural world*. In fact, this point is underscored by Cartwright's own illustration of her view (which doubles as the cover of her *The Dappled World*): she pictures the different disciplines of science as a loose collection of flexible, yet bounded balloons in no particular order; but the balloons are each tied "to the same material world" (Cartwright, 1999, 6), about which a great many systematically integrated truths *can* be formulated in plain, everyday language (e.g. "There are two trees underneath the balloons;" "The trees are to the left of a stop sign"). There is no direct analogue to this underlying, unified material world for the normative truths of ethics and politics.

This last point was driven home by David Wiggins and Bernard Williams in a little-known joint essay which is worth quoting at length:

> In science, theorists hope to find a few principles from which everything else will be deducible. … But in the case of moral philosophy what defines the subject is a highly heterogeneous set of human concerns, many of them at odds with many others of them, many of them incommensurable with many others of them. In this case there is no reason to think that what is needed is a theory to discover *underlying order*. This is not a subject after all where very much is hidden. Or rather what is hidden is hidden in a psychological or interpretive sense. There is no question of a secret axiological ordering principle. There is no deeper level of reality comparable to the microscopic or submicroscopic level explored by chemistry and physics which it is the moral philosopher's duty to probe. And where one can make no sense of there being such a level, the idea, urged by some moral philosophers, of finding the "simplest theory" which will "save the phenomena" (in the normal acceptation of the phrase) is nearly meaningless. In a physical subject matter where people speak in this way, the word "simplest" can in the respectable case (where theory is more than a mere curvefitting exercise) be provided with an independent (but subjectbound) elucidation; and the equations which yield the required curve on a graph can then be thought of as homing, as it were, upon an independent physical reality. In the moral theory case it seems perverse in the extreme to look for anything like this—unless we think of the theorist's principles as homing upon a mental reality, viz. the moral and valuational consciousness from which the whole construction originates. But that is not something hidden or unobservable. It can be consulted at any time and it may be that it can be improved: but not by the regimentation of a theory whose sole claim to authority resides in

---

[12] Thus, Cartwright herself concedes that there are "pockets of precise order" (1999, 57), as does Gaukroger (2020, 10) even as he warns against overgeneralizing from localized systematicity.

its fallacious title to express the underlying and hidden laws of that consciousness. (Wiggins & Williams, 1978, xxxviii–xxxix)

The thought that there is no *hidden underlying order* to be discovered in normative domains has been invoked to argue against philosophy's aspiration to organize normative truths into systematic ethical theories—most sustainedly perhaps by Sophie Grace Chappell, who also emphasizes that embracing asystematicity is fully compatible with realism about values.[13] However, the crucial point for present purposes is the prior one about the *structure* of normative truths: the vast landscape of ground-level normative truths that both philosophers and LLMs seek to capture, either by compressing them into a tidy theory or by compressing them into model weights, *itself* lacks structural cohesion. Inconsistency and incoherence do not emerge only at the theoretical level at which the implications of one theoretical edifice conflict or fail to cohere with the implications of another theoretical edifice (which would mirror the way in which scientific models relate according to Cartwright). Rather, inconsistency and incoherence extend all the way down to the ground level of the very values that are being modelled.

These conflicting ground-level truths can take one of two forms.[14] In the first case, one and the same action appears to be one I *ought* to perform in view of some of its features and at the same time appears to be one I ought *not* to perform in view of some of its other features: in light of value *x*, I ought to $\phi$; but in light of value *y*, I ought not to $\phi$. Deciding whether or not to $\phi$ then requires one to judge the relative importance of the features that count for and against the action in this situation.

In the second case, there are two actions I each ought to perform, but I cannot perform both: in light of value *x*, I ought to $\phi$; but in light of value *y*, I ought to $\psi$ instead, and I cannot do both. The fact that I cannot do both may be due to a contingent empirical feature of the world—jackhammers working as they do at present, it may be impossible to be a jackhammer operator while moonlighting as a concert pianist.[15] But equally, the conflict may be inherent to the values themselves. There may be an inherent tension between *x* and *y*—liberty and equality, say, or truthfulness and happiness, or security and privacy—in that the sustained realization of *x* can only happen at the expense of the sustained realization of *y*, and vice versa. Again, a judgement of importance seems to be required to decide which value should prevail in a given situation, and how far one should go in sacrificing the realization of the other value to that end.

Pluralists maintain that such conflicting truths cannot always be analysed away; the conflict between them may be genuine and irreducible. This highlights a stark asymmetry between systematic and asystematic truth. When one discovers a conflict between two beliefs about a systematic domain, the discovery of the conflict is normally taken as evidence of some *epistemic error*, and one's confidence in the conflicting beliefs is *weakened* until the error is found and at least one of the two offending beliefs is abandoned altogether. But in asystematic domains, the discovery of conflict need not be evidence of epistemic error, and one's confidence in the conflicting beliefs need not be weakened at all. Rather, one's discovery of the conflict makes one realize that one faces a dilemma, or at least a trade-off, and that this requires a judgement about what is more important in that particular situation. And once that judgment has been made and one belief has prevailed over the other, the overruled belief does not disappear. Instead, it now registers as *regret at the real costs incurred* in terms of the value that was overruled in the name of some other value that struck one as more important in that situation.[16] Both conflicting beliefs thus endure—though after one prevails, the other resurfaces in a different guise: as regret, or as a sense of loss, which acknowledges the reality and force of the consideration that was not acted upon, and which may subsequently motivate further action, such as showing remorse, making amends, issuing an apology, or offering some sort of compensation or reparation.[17]

---

[13] See not only the books published under the name Sophie Grace Chappell (2015a, 2015b, 2022), but also those published under the name Timothy Chappell (2009; 2015a, 2015b).

[14] I draw here on Williams's (1973, 171) discussion of conflicts of ought.

[15] I take the example from Millgram and Thagard (1996, 73).

These ineliminable conflicts between truths about values imply that, at the end of the line, there may be no systematic harmony to be had in normative domains: the various normative truths expressing our values and describing what the world ought to be like do not fit neatly into a unified, consistent, and coherent system. Truths about values, we might say, are at least partly *asystematic*. We constantly acknowledge this asystematicity when, by showing regret or remorse, we acknowledge that a real loss was incurred as a result of our doing something we nevertheless *had* to do. If all truths about normative domains fit together into a harmonious whole, we could realize all our values without painful trade-offs or remainders. To acknowledge that we cannot do so is to acknowledge that these truths are to some extent asystematic.

The point can be vividly put in terms of the distinction between a map and the landscape it depicts. Truths in certain domains, such as geography, form a systematically integrated whole, because the landscape they map itself forms a systematically integrated whole—in this case, the Earth, whose *ontological* systematicity ensures the *logical* systematicity of geographical truths such as "Paris is west of Prague" and "Prague is east of Paris." But if pluralists are correct, then truths in other domains, such as normative truths about ethics and politics, do not form a systematically integrated whole, because the landscape they map itself does not form a systematically integrated whole; rather, it forms a fragmented, tension-ridden, disparate, and disconnected landscape. This should come as no surprise if we regard the normative landscape as the historical deposit of a variety of influences and vastly different traditions.[18] Why should we expect the vicissitudes of cultural history, with all the disparate traditions of normative reflection they jumbled together and continually reconfigured in often contingent and messy ways, to produce, of all things, a practice of normative reflection tracking a neatly integrated normative landscape? Such a process is far more likely to have produced a disconnected patchwork of conflicting normative considerations. Those who insist that the normative landscape itself already exhibits perfect systematicity, pluralists maintain, owe us an explanation of how that systematicity is supposed to have got there.[19]

---

[16] These features of value conflicts are discussed in detail in Williams (1973) and Queloz (2024b).

[17] On regret or a sense of loss as an acknowledgement of genuine conflicts of values, see Williams (1973, 1981a, b, 2005a), Queloz (2024a), and Cueni (2024).

[18] See MacIntyre (2007) and Williams (2005b, 136–37).

Insofar as truths about values are asystematic, the proper orientation towards them is not to shoehorn them into a unified and coherent system, but to aim at a nuanced understanding of the subtle interplay and trade-offs between them. As Nora Hämäläinen emphasizes, striving for systematicity and coherence may not be the "proper orientation in the moral realm," because "the gaps and leaps in our moral vocabularies and frameworks may be essential to the object of investigation—morality—rather than faults in our understanding of it, that need to be corrected by a more coherent, unitary perspective" (2009, 548). Our evaluative beliefs are accordingly free to be as irreducibly disparate, inconsistent, and tension-ridden as our values themselves are—indeed, our evaluative beliefs *should* mirror this asystematicity if they are to be true to our values.

The asystematicity of normative truths in turn has implications for the prospects of LLMs. LLMs excel at identifying and leveraging patterns and are becoming increasingly good at making inferences within systematic domains. But any machine learning approach that relies on identifying and leveraging systematic and consistent patterns will have a harder time modelling a normative landscape that lacks systematicity and consistency. If pluralists are right and normative domains are at least partly asystematic, then attempts to model human values cannot expect to receive the kind of support from the systematicity of truth that they can in principle rely on in systematic domains such as geography. If normative truths are asystematic, these truths will not exhibit the same degree of inferential interconnectedness and redundancy that geographical truths exhibit. When confronted with the inherent and irresoluble conflicts that value pluralism highlights, therefore, LLMs' ability to extrapolate from incomplete training data and comprehensively model human values will be hampered. If pluralists are right, the asystematicity of truth in normative domains is a significant hurdle for AI models.

This is a different hurdle from the one that Sorensen et al. (2024) have recently identified. Their worry is that insofar as AI systems are statistical learners that aggregate vast amounts of data and fit it to averages, they are ill-poised to learn about inherently conflicting values, because they risk "washing out" (Sorensen et al., 2024, 19937) just the value conflicts we want them to model. The same "washing out" problem also afflicts the attempt of Feng et al. (2024) to extract normative requirements from their LLM's pre-training data: they filter the normative requirements to ensure their consistency. But if the normative truths that LLMs are supposed to model are themselves inconsistent, this filtering process effectively distorts the model's grasp of the normative landscape it is trying to map. When dealing with asystematic domains, the very strategy that promises to help LLMs self-complete and self-correct in mapping out systematically integrated domains thus turns into a counterproductive strategy that risks distorting the map.

---

[19] See Williams (1995c, 189).

As Sorensen et al. (2024) show, a significant step towards overcoming this difficulty is to accommodate value conflicts by explicitly representing them within a dataset such as ValuePrism. This expressly "value pluralist" dataset leverages GPT-4's open-text generative capabilities to make explicit the wide variety of human values that are encoded in its pretraining data. The resulting dataset purports to cover 218 k examples of values, rights, and duties contextualized in terms of 31 k human-described situations (obtained by filtering 1.3 M human-written situations sourced from the Allen Institute's Delphi demo).

Trained on such a data set, models such as Value Kaleidoskope (Kaleido) manage to explicitly represent conflicts between values (Sorensen et al,. 2024). Given a description of a situation (e.g. "Telling a lie to protect a friend's feelings"), Kaleido begins by exploratively generating one-hundred normative considerations (e.g. "Honesty," "Friend's well-being") before filtering them according to their relevance to the situation. It then removes repetitive items based on textual similarity, and computes relevance and valence scores for each of the remaining normative considerations (where the relevance is some number between 0 and 1, and the valence is *support*, *oppose*, or *either, depending on context*). Finally, it generates a post-hoc rationale explaining why each of the normative considerations bears on the situation (e.g. "If you value honesty, it may be better to tell the truth even if it hurts feelings").

What an AI model along these lines *can* do is to explicitly acknowledge and advert to the conflicting values implicit in its training data. This can be a valuable form of assistance, especially if it reminds one of the *variety* of values that bear on the appraisal of a situation. Even now, LLMs are much more effective than humans at exploratively overgenerating potentially relevant considerations that can then be filtered by relevance. This form of assistance is apt to draw one's attention to relevant aspects one had not yet thought of considering.

But what even a pluralist AI model such as Kaleido *cannot* do is to overcome the limitation imposed by the asystematicity of normative truth on its capacity to move beyond its training data. On the pluralist picture, even an LLM trained to acknowledge the reality of value conflicts will not be able to overcome omissions and inaccuracies in the training data *by leveraging the systematicity of truth* any more than the GPT-4 model it is based on. Insofar as the truth in normative domains is asystematic, this will rob both kinds of models of their capacity to rely on the consistency and coherence of the truth to work towards comprehensiveness.

## 5 The Less Systematicity, the More Human Agency

What are the implications of the asystematicity of truth for AI's potential as a moral advisor, which is garnering increasing attention now that LLMs rival professional ethicists in perceived moral expertise?[20] In this final section, I shall argue that the less AI can rely on the systematicity of truth, the less we can rely on AI to do our practical deliberation for us. This is because the less the truth in normative domains is systematic, the more of a role there is for human agency and individuality in practical deliberation.

---

[20] Using a "Moral Turing Test," researchers found that people perceive practical advice from GPT-4o as more moral, trustworthy, thoughtful, and correct than that of professional ethicists in the New York Times column "The Ethicist" (Dillion et al., 2025). See, e.g., Giubilini and Savulescu (2018); Constantinescu et al. (2021); Landes, Voinea, and Uszkai (2024) for discussions of AI's potential as a moral advisor.

To see this, consider how the contrast between the systematicity of empirical domains on the one hand and the asystematicity of normative domains on the other produces a corresponding contrast in the structure of theoretical and practical deliberation (i.e. deliberation about what to believe and deliberation about what to do).

When deliberating about what to believe about some systematic domain such as geography, my belief-formation aims at a set of truths that are consistent and coherent, and what I end up truly believing must be consistent and cohere with what others end up truly believing. In other words, the systematicity of truth makes it the case that what *I* should believe is the same as what *anyone* should believe. The conclusion that *I* should believe that Paris is west of Prague then feels *derivative*, following as it does from a more general truth, namely that *anyone* should believe that Paris is west of Prague. In that sense, the deliberation is only *incidentally mine*.

When deliberating about what I should *do*, by contrast, the equivalence between what *I* should do and what *anyone* should do breaks down.[21] The more normative truths conflict, presenting us with considerations that pull in different directions while remaining irreducibly distinct, incompatible, and incommensurable, the more *judgements of importance* will be required to determine which consideration should prevail over which in a given situation of practical deliberation.

These judgements of importance cannot be outsourced to an algorithm. One might see an attempt to do so in the "relevance scores" that Kaleido ascribes to values as a way of assessing to what extent they bear on a situation. But these relevance scores remain crucially different from the judgements of importance I have in mind. As the developers of Kaleido note, their model's training data is *synthetic*, i.e. itself generated by an AI model—in this case, GPT-4. This means that Kaleido is not directly trained to predict whether *humans* would find a value relevant; rather, "the model's training objective is in fact closer to predicting whether a given value was likely to be generated for a particular situation by GPT-4" (Sorensen et al., 2024, 19942). Consequently, the relevance scores generated by the model capture no more than the likelihood of a certain text string figuring in text generated by GPT-4 in response to the description of a situation. The relevance score thus measures the *statistical* relevance of a type of consideration to a type of situation. But a particular consideration's importance to the agent in a particular situation goes significantly beyond such merely statistical relevance.

One might think that the problem with statistical relevance is simply that it fails to be normative, since what we really want to know is which considerations are *normatively* relevant to the situation at hand. This is true as far as it goes, but the point about importance goes further. Normative relevance is something that a more sophisticated, pluralist AI model could in principle approximate by becoming a reliable predictor of what humans would deem normatively relevant. The crucial question of *which* humans could be addressed by fine-tuning the model on the individual user's values, as Giubilini and Savulescu (2018), Constantinescu et al. (2021), and Landes, Voinea, and Uszkai (2024) have suggested, thereby enabling the model to mirrors its users' judgements of normative relevance back at them and help them be more consistent and coherent in their own practical deliberation.

---

[21] As pointed out notably by Williams (1985, 76–77; 1995a, 123–125; 1995b, 170), whose argument I develop and apply to the contrast between systematic and asystematic domains here.

However, even an AI model that reliably tracked normative relevance to the user could not thereby close the gap to what is important *to the agent in a particular situation*. That judgement is one that necessarily falls to the agent, and cannot be offloaded onto anyone else.

This fact is obscured by an ambiguity in the question "What should I do?", which produces the impression that I might have the question answered on my behalf by someone else, or even by an AI model acting as a moral advisor. But we have to distinguish two kinds of "should": the "should" that figures in the *impersonal* "What should I do?"-question and the "should" that figures in the *first-personal* "What should I do?"-question.[22] The impersonal "What should I do?"-question coincides with the "What is to be done?"-question, which asks for the recommendation of a course of action in light of all the normatively relevant considerations. But even if considering all the normatively relevant considerations yields a clear answer to the impersonal "What should I do?"-question, there remains a *further* question for the agent who is to act on that answer: what, given all that, should *I* now do in this particular situation? This is not simply a repetition of the earlier question. And even if my answer to that further question coincides with the answer to the former question, this will not simply be a repetition of the answer. It will be an expression of the agent's judgement that what normative considerations suggest is to be done is indeed what he or she should do. This becomes evident if we consider a case in which normative considerations clearly indicate that I should $\phi$, but I would very much like to $\psi$, and I cannot do both.[23] If I then ask myself: "What should I do?", I am not asking what course of action is favoured by the relevant normative considerations. I already know that. I am asking myself whether I really should $\phi$, as the relevant normative considerations suggest I should, or whether I should follow my inclination to $\psi$ instead.

We must accordingly be careful not to identify the "should" that figures in statements of what course of action normatively relevant considerations recommend with the "should" in which the agent's own practical deliberation must ultimately issue. The former is only incidentally first-personal, and can equally well be answered in the third person ("What he should do is …"). The latter, however, is essentially first-personal, and can only be answered in the first person.

---

[22] Here, I draw out the consequences for AI of Williams's suggestion that practical thought is "radically first-personal" (Williams 1985, 23). See also Queloz (2021).

[23] This adapts an argument offered in Williams (1973, 183–185) to distinguish two kinds of *ought*; see also Williams (1995a, 123–125).

Suppose a sophisticated LLM trained to track what a human agent *A* deems normatively relevant is used to answer a practical question. If the question really is a practical question, i.e. a question about what to *do*, *A* will need to decide whether to *enact* the answer. For *A* to enact the answer, however, it is not sufficient for the AI model to think—or for its textual output to assert—that *A* should $\phi$. *A herself* needs to conclude that she should $\phi$, in the irreducibly first-personal sense of "should." This requires that $\phi$-ing should make sense *to A* in terms of *her* judgement of what is most important to her in this situation. The practical question of what *A* should do, even if competently answered by the AI model in terms of all the normatively relevant considerations that have carried weight with *A* in the past, still culminates, at the end of the line, in a first-personal question for *A* that can only be answered in terms of *A*'s own judgements of importance in the particular case.

Might *A* not leapfrog this first-personal question by making it an overriding principle of hers to enact whatever the AI model identifies as the thing to be done? This would be tempting if the model had a track record of providing sensible advice. But notice that even then, *A* would still be forced to answer first-personal practical questions herself. If, for instance, she had the thought "The AI, who has a good track record on this kind of question, said that $\phi$-ing was the thing to be done," she would still be bound to confront questions such as "Should I $\phi$ *now* or *later*?" and "Should I $\phi$ *in this way* or *in that way*?" Practical deliberation has an irreducibly first-personal dimension that expresses itself in the enduring availability of this further sense of "should." And parallel arguments can be run to distinguish two kinds of "shall" and two kinds of "ought" (as in "What shall I do?" and "What ought I to do?").[24]

Of course, if this is right, then there would be a first-personal version of the "What should I do?"-question even if all normative truths could be systematically integrated into a harmonious whole.[25] By increasing the conflictual character of practical deliberation, however, the asystematicity of normative domains *accentuates and amplifies* the role of the agent. The more irreducible conflicts between incommensurable values the normative landscape confronts us with, the more we will face uncomfortable binds, true dilemmas, and tragic choices, and the more first-personal judgements about what is most important in a given situation will be required.

The pluralist picture thereby not only renders the radically first-personal character of practical deliberation *salient* where the Kantian or utilitarian pictures *occlude* it by suggesting that there are no truly irreducible value conflicts for the individual to resolve; the pluralist picture also gives a more active role to individual agents by requiring them to determine how these value conflicts are to be navigated.

We might put this by saying that while all practical deliberation is to some degree first-personal regardless of the systematicity of truth, the asystematicity of the truth in normative domains *adds* to this first-personal dimension by allocating a *greater* role to the agent's judgements of importance. For, in navigating a conflictual normative landscape, I am forced to rely to a greater extent on my judgements of what strikes me as most important in a particular situation.

---

[24] See Williams (1973, 183–185; 1995a, 123–125).

[25] Similarly, even questions about systematically integrated non-normative domains—such as: "Did Keats die in Rome?"—have first-personal analogues, such as: "What should I believe about whether Keats died in Rome?" But these are first-personal only incidentally. One could equally well ask: "What should anyone believe about whether Keats died in Rome?".

This comes out well in what Ruth Chang has called the "hard choices" (2017) that turn out to be ubiquitous once we recognize that normativity is *tetrachotomous* rather than trichotomous in structure—two items can be normatively related in *four* rather than three different ways: (i) one can be *better* than the other; (ii) one can be *worse* than the other; (iii) they can be *equally good*, so that one may as well flip a coin; and (iv) they can be *on a par*, i.e. incommensurable, yet in the same neighbourhood of value. Hard choices are choices between such options that are on a par, and remain on a par even once all the relevant information is in. One choice is better in some respects, while the other is better in other respects. This does not mean that we might as well flip a coin, however. The decision can still be rational in the sense of being grounded in reasons rather than arbitrary. But coming to a rational decision requires the agent to go beyond the passive role of registering the relevance of independently given normative considerations. The agent has to play a more active role in the decision and consider which aspects are more important *to him or her*. In reminding us that even commonplace instances of practical deliberation are essentially dependent on input from the agent whose action is at issue, hard choices encourage "a fundamental shift in our understanding of what it is to be a rational agent, one that puts active, creative human agency at the center of rational thought and action" (Chang, 2023, 173).

Imagine, for instance, that I am deliberating over whether I should pursue a career in philosophy or a career in consulting. Even once all the relevant normative considerations for and against each of these career choices have been exhaustively listed and carefully considered, there remains the question of what *I* should do, especially if the various considerations do not all harmoniously assemble into an unequivocal answer. Chang presents such choices in markedly voluntaristic terms, as answerable primarily to the agent's will—on her account, it is by being willing to *commit* one way or the other that I *create* the reasons that make the choice rational.[26]

But one can also think of the process of determining what one should do as having the character of a *discovery* about oneself—and one, crucially, that only the agent him- or herself can make. Coming to a decision forces me to ask not just which considerations strike me as more important, but which considerations are more important *to me*. The decision is not merely *first*-personal in the way that every practical decision ultimately must be, but, as we naturally put it, *personal*. In coming to the conclusion that *I* should pursue a career in philosophy, because certain considerations favouring that choice are particularly important to me, I then express something distinctive of myself—something which may already have been fully formed before the process of deliberation, or which may have assumed a determinate form only through that process, but which nevertheless presents itself to me not as an expression of my will, but as a discovery about myself. Though the decision should of course still be informed by the relevant impersonal normative considerations, it should not *just* be responsive to them, but should also be true to who I am, or discover myself to have become. We might say that the form of truthfulness involved is two-faced: it encompasses being true to oneself as well as being true to the normative facts. In other words, the decision involves a demand for authenticity as well as for responsiveness to impersonal reasons.

---

[26] See Chang (2002, 2009, 2013, 2016). Drawing on Chang's work, Goodman (2021) argues that the existence of hard choices imposes limits on how much practical reasoning AI models can do on our behalf.

As a result of this demand for authenticity, the conclusion that *I* should pursue a career in philosophy does not feel derivative, because it does not follow from the more general thought that *anyone* should pursue a career in philosophy. The deliberation is not just incidentally, but *essentially mine*. Practical deliberation from the point of view of a quasi-omniscient AI cannot be a substitute for this.[27] As Williams remarks: "my life, my action, is quite irreducibly mine, and to require that it is at best a *derivative* conclusion that it should be lived from the perspective that happens to be mine is an extraordinary misunderstanding" (1995b, 170). Far from being answerable only to impersonal normative considerations that an AI might weigh against each other just as well or even better than a human agent, practical deliberation systematically possesses a *first-personal* dimension, and sometimes even a *personal* dimension, in virtue of which the decision cannot be outsourced to anything or anyone else, but is essentially the agent's own.

AI models built on the assumption that practical reasoning is impersonal, and that the question "What should I do?" is equivalent to the question "What should anyone do?," neglect these first-personal and personal dimensions of practical deliberation. This remains true even if, like Kaleido, a model takes the plurality and incompatibility of values into account. For its conclusion will still take an impersonal form: it will state that, in a situation in which such-and-such conflicting values are likely to be relevant, $\phi$-ing is the thing to do. If we are to be true to the first-personal and personal nature of practical thought, however, this can at most be advisory *input* to the agent's deliberation. The conclusion as to what the agent should actually do still has to be reached *by the agent whose practical decision it is*. This vindicates the pluralist conviction that "practical decision could not in principle be made completely algorithmic, and … a conception of practical reason which aims at an algorithmic ideal must be mistaken" (Berlin & Williams, 1994, 307).

The upshot is that the less systematicity normative domains exhibit, the less AI can rely on the systematicity of truth, and the less we can rely on AI to do our practical deliberation for us. There is correspondingly more of a role for human agency and individuality in navigating conflicts of values and making hard choices. The less systematicity, the more human agency.

---

[27] For a complementary argument why we have reasons not to want an AI that lets one know who one is and what one should do, see Leuenberger (2024).

# 6 Conclusion

Insofar as normative domains exhibit asystematicity, Amodei's optimism thus looks ill-founded: LLMs cannot rely on truths forming a systematic web across the board to self-complete and self-correct. Alternative ways for LLMs to move beyond their training data may yet emerge. But if the pluralist picture of normative domains is correct, LLMs cannot leverage the systematic harmony of these domains. Accordingly, it should to that extent be harder for LLMs to comprehensively model normative domains.

What is more, that lack of systematicity itself calls for human beings to stay in the decision-making loop. The more asystematic normative domains are, the more judgements of importance by the agent are required, and these judgements of importance express and underscore first-personal and personal dimensions of practical deliberation that cannot be outsourced to AI models. Sometimes, the point is not that a decision should be absolutely and objectively the best one, but that it should be *ours*.

**Acknowledgements** I am grateful to the Editor as well as to two anonymous reviewers for their valuable comments on this paper. Pierre Beckmann, Markus Stepanians, and Mümün Gencoglu provided insightful feedback on an earlier draft. I am also grateful to Ben Matheson, Constant Bonard, and the participants of the colloquium on issues in practical philosophy at the University of Bern.

**Authors' Contributions** Not applicable.

**Funding** Open access funding provided by University of Bern. Funded by the Swiss National Science Foundation.

**Data Availability** Not applicable.

## Declarations

**Ethics Approval and Consent to Participate** Not applicable.

**Consent for Publication** Not applicable.

**Competing interests** The author declares that he has no competing interests.

## References

Abela, P. (2006). The Demands of Systematicity: Rational Judgment and the Structure of Nature. In G. Bird (Ed.), *A Companion to Kant* (pp. 408–422). Blackwell.

Amodei, D. (2024). "What if Dario Amodei Is Right About A.I.?." Interview by Ezra Klein. *The Ezra Klein Show*, New York Times Opinion, April 12, 2024. https://www.nytimes.com/2024/04/12/opinion/ezra-klein-podcast-dario-amodei.html

Askell, A., Bai, Y, Chen, A., Drain, D, Ganguli, D, Henighan, T., Jones, A., Joseph, N., Mann, B, & Das-Sarma, Nova. (2021). "A general language assistant as a laboratory for alignment." *arXiv preprint* arXiv:2112.00861.

Bai, Y., Kadavath, S., Kundu, S., Askell, A., Kernion, J., Jones, A., ... & Kaplan, J. (2022). Constitutional ai: Harmlessness from ai feedback. arXiv preprint arXiv:2212.08073.

Beisbart, C. (forthcoming-a). "Epistemology of Artificial Intelligence." In *The Stanford Encyclopedia of Philosophy*. Edited by Edward N. Zalta.

Beisbart, C. (forthcoming-b). "In Which Ways Is Machine Learning Opaque?." In *Philosophy of Science for Machine Learning: Core Issues and New Perspectives*. Edited by Juan Durán and Giorgia Pozzi. Dordrecht: Springer.

Berlin, I. (2002). Two Concepts of Liberty. In H. Hardy (Ed.), *Liberty* (pp. 166–217). Oxford University Press.

Berlin, I. (2013). The Pursuit of the Ideal. In H. Hardy (Ed.), *The Crooked Timber of Humanity: Chapters in the History of Ideas* (pp. 1–20). Princeton University Press.

Berlin, I., & Williams, B. (1994). Pluralism and Liberalism: A Reply. *Political Studies, 42*(2), 306–309.

Blum, C. (2023). Value Pluralism versus Value Monism. *Acta Analytica, 38*(4), 627–652.

Brooks, T., & Stein, S. (Eds.). (2017). *Hegel's Political Philosophy: On the Normative Significance of Method and System*. Oxford University Press.

Buckner, C. (2023). *From deep learning to rational machines: What the history of philosophy can teach us about the future of artificial intelligence*. Oxford University Press.

Cartwright, N. (1983). *How the Laws of Physics Lie*. Oxford University Press.

Cartwright, N. (1999). *The Dappled World: A Study of the Boundaries of Science*. Cambridge University Press.

Chang, R. (Ed.). (1997a). *Incommensurability, Incomparability, and Practical Reason*. Harvard University Press.

Chang, R. (1997b). Introduction. In R. Chang (Ed.), *Incommensurability, Incomparability, and Practical Reason* (pp. 1–34). Harvard University Press.

Chang, R. (2002). *Making Comparisons Count*. Routledge.

Chang, R. (2009). Voluntarist Reasons and the Sources of Normativity. In D. Sobel & S. Wall (Eds.), *Reasons for Action* (pp. 243–271). Cambridge University Press.

Chang, R. (2013). Commitments, Reasons, and the Will. In R. Shafer-Landau (Ed.), *Oxford Studies in Metaethics* (Vol. 8, pp. 74–113). Oxford University Press.

Chang, R. (2015a). Value Incomparability and Incommensurability. In I. Hirose & J. Olson (Eds.), *The Oxford Handbook of Value Theory* (pp. 205–224). Oxford University Press.

Chang, R. (2015). Value Pluralism. *In International Encyclopedia of the Social & Behavioral Sciences, 25*, 21–26.

Chang, R. (2016). Comparativism: The Grounds of Rational Choice. In E. Lord & B. Maguire (Eds.), *Weighing Reasons* (pp. 213–240). Oxford University Press.

Chang, R. (2017). Hard Choices. *Journal of the American Philosophical Association, 3*(1), 1–21.

Chang, R. (2023). Three Dogmas of Normativity. *Journal of Applied Philosophy, 40*(2), 173–204.

Chappell, T. (2009). *Ethics and Experience: Life Beyond Moral Theory*. Durham: Acumen.

Chappell, S. G. (Ed.). (2015a). *Intuition, Theory, and Anti-Theory in Ethics*. Oxford University Press.

Chappell, S. G. (2022). *Epiphanies: An Ethics of Experience*. Oxford University Press.

Chappell, T. (2015b). *Knowing What To Do: Imagination, Virtue, and Platonism in Ethics*. Oxford University Press.

Constantinescu, M., Vică, C., Uszkai, R., & Voinea, C. (2021). Blame It on the AI? On the Moral Responsibility of Artificial Moral Advisors. *Philosophy and Technology, 35*(2), 1–26.

Cueni, D. (2024). Constructing Liberty and Equality – Political, Not Juridical. *Jurisprudence, 15*(3), 341–360.

Cummins, R., Blackmon, J., Byrd, D., Poirier, P., Roth, M., & Schwarz, G. (2001). Systematicity and the Cognition of Structured Domains. *Journal of Philosophy, 98*(4), 167–185.
Dancy, J. (1995). In Defense of Thick Concepts. *Midwest Studies in Philosophy, 20*(1), 263–279.
Dancy, J. (2004). *Ethics without Principles*. Clarendon Press.
Dillion, D., Mondal, D., Tandon, N., & Gray, K. (2025). AI language model rivals expert ethicist in perceived moral expertise. *Scientific Reports, 15*(1), 4084. https://doi.org/10.1038/s41598-025-86510-0
Downes, S. M., Forber, P., & Grzankowski, Alex. (2024). "LLMs are Not Just Next Token Predictors." *arXiv* arXiv:2408.04666.
Dupré, J. (1995). *The Disorder of Things: Metaphysical Foundations of the Disunity of Science*. Harvard University Press.
Elazar, Y., Kassner, N., Ravfogel, S., Ravichander, A., Hovy, E., Schütze, H., & Goldberg, Y. (2021). Measuring and Improving Consistency in Pretrained Language Models. *Transactions of the Association for Computational Linguistics, 9*, 1012–1031.
Feng, N., Marsso, L., Yaman, S. G., Standen, I., Baatartogtokh, Y., Ayad, R., ... & Chechik, M. (2024, June). Normative requirements operationalization with large language models. In 2024 IEEE 32nd International Requirements Engineering Conference (RE) (pp. 129-141). IEEE.
Feyerabend, P. (1993). *Against Method* (3rd ed.). Verso.
Franks, P. W. (2005). *All or Nothing: Systematicity, Transcendental Arguments, and Skepticism in German Idealism*. Harvard University Press.
Galison, P., & Stump, D. J. (Eds.). (1996). *The Disunity of Science: Boundaries, Contexts, and Power*. Stanford University Press.
Gaukroger, S. (2020). *Civilization and the Culture of Science: Science and the Shaping of Modernity, 1795–1935, Civilization and the Culture of Science*. Oxford University Press.
Giubilini, A., & Savulescu, J. (2018). The Artificial Moral Advisor. The "Ideal Observer" Meets Artificial Intelligence. *Philosophy and Technology, 31*(2), 169–188.
Goldstein, S., & Levinstein, B. A. (2024). Does ChatGPT Have a Mind?. arXiv preprint arXiv:2407.11015.
Goodman, B. (2021). "Hard Choices and Hard Limits for Artificial Intelligence." Proceedings of the 2021 AAAI/ACM Conference on AI, Ethics, and Society.
Greene, J. (2013). *Moral Tribes: Emotion, Reason, and the Gap Between Us and Them*. Penguin Press.
Greene, J. D. (2016). Our Driverless Dilemma. *Science, 352*(6293), 1514–1515.
Guyer, P. (2003). Kant on the Systematicity of Nature: Two Puzzles. *History of Philosophy Quarterly, 20*(3), 277–295.
Guyer, P. (2005). *Kant's System of Nature and Freedom*. Oxford University Press.
Hämäläinen, N. (2009). Is Moral Theory Harmful in Practice?—Relocating Anti-theory in Contemporary Ethics. *Ethical Theory and Moral Practice, 12*(5), 539–553.
Heathwood, C. (2015). Monism and Pluralism about Value. In I. Hirose & J. Olson (Eds.), *The Oxford Handbook of Value Theory* (pp. 136–157). Oxford University Press.
Herrmann, D. A., & Levinstein, B. A. (2025). Standards for Belief Representations in LLMs. *Minds and Machines, 35*(1), 1–25.
Hooker, B., & Little, M. O. (Eds.). (2000). *Moral Particularism*. Oxford University Press.
Jang, M., Lukasiewicz, T. (2023). "Consistency Analysis of ChatGPT." Singapore, December.
Kambartel, F. (1969). ""System" und "Begründung" als wissenschaftliche und philosophische Ordnungsbegriffe bei und vor Kant." In *Philosophie und Rechtswissenschaft: Zum Problem ihrer Beziehung im 19. Jahrhundert*. Edited by Jürgen Blühdorn and Joachim Ritter, 99–122. Frankfurt am Main: Klostermann.
Kekes, J. (1993). *The Morality of Pluralism*. Princeton University Press.
Kitcher, P. (1986). Projecting the Order of Nature. In R. Butts (Ed.), *Kant's Philosophy of Material Nature* (pp. 201–235). D. Reidel.
Kretzmann, Norman, & Eleonore Stump. (1989). *The Cambridge Translations of Medieval Philosophical Texts: Volume 1, Logic and the Philosophy of Language*. Cambridge: Cambridge University Press.
Kumar, Ashutosh, & Aditya Joshi. (2022). "Striking a Balance: Alleviating Inconsistency in Pre-trained Models for Symmetric Classification Tasks." Dublin, Ireland, May.
Landes, E., Voinea, C., & Uszkai, R. (2024). Rage against the authority machines: how to design artificial moral advisors for moral enhancement. AI and SOCIETY, 1-12.
Larmore, C. (1987). *Patterns of Moral Complexity*. Cambridge University Press.

Lee, H., Phatale, S., Mansoor, H., Lu, K. R., Mesnard, T., Ferret, J., ... & Rastogi, A. (2023). Rlaif: Scaling reinforcement learning from human feedback with ai feedback.

Leuenberger, M. (2024). Should You Let AI Tell You Who You Are and What You Should Do? In D. Edmonds (Ed.), *AI Morality* (pp. 160–169). Oxford University Press.

Levinstein, B. A., & Herrmann, D. A. (2024). "Still No Lie Detector for Language Models: Probing Empirical and Conceptual Roadblocks." *Philosophical Studies*, 1–27.

Liu, Y., Guo, Z., Liang, T., Shareghi, E., Vulić, I., & Collier, N. (2024). "Measuring, evaluating and improving logical consistency in large language models." *arXiv* arXiv:2410.02205.

Losano, M. G. (1968). *Sistema e struttura nel diritto, vol. 1: Dalle origini alla scuola storica*. Turin: Giuffrè.

MacIntyre, A. C. (2007). *After Virtue: A Study in Moral Theory* (3rd ed.). University of Notre Dame Press.

Mason, E. (2023). "Value Pluralism." In *The Stanford Encyclopedia of Philosophy*. Edited by Edward N. Zalta. Summer 2023 ed.

Messer, A. (1907). Besprechung von Otto Ritschl: System und systematische Methode in der Geschichte des wissenschaftlichen Sprachgebrauchs und der philosophischen Methodologie. *Göttinger Gelehrte Anzeigen, 169*(8), 659–666.

Millgram, E., & Thagard, P. (1996). Deliberative Coherence. *Synthese, 108*(1), 63–88.

Nagel, T. (2001). Pluralism and Coherence. In M. Lilla, R. Dworkin, & R. Silvers (Eds.), *The Legacy of Isaiah Berlin* (pp. 105–111). New York Review of Books.

Neurath, O. (1935). Einheit der Wissenschaft als Aufgabe. *Erkenntnis, 5*, 16–22.

Ouyang, L., Jeffrey, Wu., Jiang, Xu., Almeida, D., Wainwright, C., Mishkin, P., Zhang, C., Agarwal, S., Slama, K., & Ray, A. (2022). Training language models to follow instructions with human feedback. *Advances in Neural Information Processing Systems, 35*, 27730–27744.

Queloz, M. (2021). Choosing values? williams contra Nietzsche. *The Philosophical Quarterly, 71*(2), 286–307. https://doi.org/10.1093/pq/pqaa026

Queloz, M. (2024a). The Dworkin-Williams debate: Liberty, conceptual integrity, and tragic conflict in politics. *Philosophy and Phenomenological Research, 109*(1), 3–29.

Queloz, M. (2024b). Moralism as a dualism in ethics and politics. *Political Philosophy, 1*(2), 433–462. https://doi.org/10.16995/pp.17532

Queloz, M. (2025). *The ethics of conceptualization: Tailoring thought and language to need*. Oxford University Press. https://doi.org/10.1093/9780198926283.001.0001

Rescher, N. (1981). "Leibniz and the Concept of a System." In *Leibniz's Metaphysics of Nature: A Group of Essays*, 29–41. Dordrecht: Springer.

Rescher, N. (1979). *Cognitive Systematization: A Systems Theoretic Approach to a Coherentist Theory of Knowledge*. Blackwell.

Rescher, N. (2000). *Kant and the Reach of Reason: Studies in Kant's Theory of Rational Systematization*. Cambridge University Press.

Rescher, N. (2005). *Cognitive Harmony: The Role of Systemic Harmony in the Constitution of Knowledge*. University of Pittsburgh Press.

Ritschl, O. (1906). *System und systematische Methode in der Geschichte des wissenschaftlichen Sprachgebrauchs und der philosophischen Methodologie*. Bonn: C. Georgi.

Russell, S. (2019). *Human Compatible: Artificial Intelligence and the Problem of Control*. Viking.

Sandis, C. (2006). Dancy Cartwright: Particularism in the philosophy of science. *Acta Analytica, 21*(2), 30–40.

Sen, A. (1981). Plural Utility. *Proceedings of the Aristotelian Society, 81*(1), 193–216.

Sorensen, T., Jiang, L., Hwang, J. D., Levine, S., Pyatkin, V., West, P., ... & Choi, Y. (2024, March). Value kaleidoscope: Engaging ai with pluralistic human values, rights, and duties. In Proceedings of the AAAI Conference on Artificial Intelligence (Vol. 38, No. 18, pp. 19937-19947).

Stein, A von der. (1968). "Der Systembegriff in seiner geschichtlichen Entwicklung." In *System und Klassifikation in Wissenschaft und Dokumentation*. Edited by Alwin Diemer, 1–13. Meisenheim am Glan: A. Hain.

Stocker, M. (1990). *Plural and Conflicting Values*. Clarendon Press.

Tasioulas, J. (2022). Artificial Intelligence, Humanistic Ethics. *Daedalus, 151*(2), 232–243.

Thompson, K. (2017). Systematicity and Normative Justification: The Method of Hegel's Philosophical Science of Right. In T. Brooks & S. Stein (Eds.), *Hegel's Political Philosophy: On the Normative Significance of Method and System* (pp. 44–66). Oxford University Press.

Troje, H. E. (1969). "Wissenschaftlichkeit und System in der Jurisprudenz des 16. Jahrhunderts." In *Philosophie und Rechtswissenschaft: Zum Problem ihrer Beziehung im 19. Jahrhundert*. Edited by Jürgen Blühdorn and Joachim Ritter, 63–88. Frankfurt am Main: Klostermann.

Vickers, P. (2013). *Understanding Inconsistent Science*. Oxford University Press.

Vieillard-Baron, J.-L. (1975). "Le concept de système de Leibniz à Condillac." In *Akten des II. Internationalen Leibniz-Kongresses Hannover, 17.-22. Juli 1972*. Edited by Kurt Müller, Heinrich Schepers and Wilhelm Totok, 97–103. Wiesbaden: F. Steiner.

Wei, J., Tay, Y., Bommasani, R., Raffel, C., Zoph, B., Borgeaud, S., ... & Fedus, W. (2022). Emergent abilities of large language models. arXiv preprint arXiv:2206.07682.

Wiggins, D., & Williams, B. (1978). "Aurel Thomas Kolnai." In *Ethics, Value and Reality: Aurel Kolnai*, xxiii–xxxix. New Brunswick, NJ: Transaction Publishers.

Williams, B. (1973). "Ethical Consistency." In *Problems of the Self*, 166–186. Cambridge: Cambridge University Press.

Williams, B. (1981a). "Conflicts of Values." In *Moral Luck*, 71–82. Cambridge: Cambridge University Press.

Williams, B. (1981b). "Moral Luck." In *Moral Luck*, 20–39. Cambridge: Cambridge University Press.

Williams, B. (1985). *Ethics and the Limits of Philosophy. Routledge* (Classics). Routledge.

Williams, B. (1995a). "Formal and Substantial Individualism." In *Making Sense of Humanity and Other Philosophical Papers 1982–1993*, 123–34. Cambridge: Cambridge University Press.

Williams, B. (1995b). "The Point of View of the Universe: Sidgwick and the Ambitions of Ethics." In *Making Sense of Humanity and Other Philosophical Papers 1982–1993*, 153–71. Cambridge: Cambridge University Press.

Williams, B. (1995c). "What Does Intuitionism Imply?." In *Making Sense of Humanity and Other Philosophical Papers 1982–1993*, 182–191. Cambridge: Cambridge University Press.

Williams, B. (2013). "Introduction." In *Concepts and Categories: Philosophical Essays*. Edited by Henry Hardy. 2nd ed, xxix–xxxix. Princeton: Princeton University Press.

Williams, B. (2001). *Morality: An Introduction to Ethics*. Cambridge University Press.

Williams, B. (2002). *Truth and truthfulness: An essay in genealogy*. Princeton University Press.

Williams, B. (2005a). From Freedom to Liberty: The Construction of a Political Value. In G. Hawthorne (Ed.), *In the Beginning Was the Deed: Realism and Moralism in Political Argument* (pp. 75–96). Princeton University Press.

Williams, B. (2005b). Pluralism, Community and Left Wittgensteinianism. In G. Hawthorne (Ed.), *In the Beginning Was the Deed: Realism and Moralism in Political Argument* (pp. 29–39). Princeton University Press.

Ypi, L. (2021). *The Architectonic of Reason: Purposiveness and Systematic Unity in Kant's Critique of Pure Reason*. Oxford University Press.

Zhao, Y., Yan, L., Sun, W., Xing, G., Wang, S., Meng, C., ... & Yin, D. (2024). Improving the robustness of large language models via consistency alignment. arXiv preprint arXiv:2403.14221.

Zhou, J., Ghaddar, A., Zhang, G., Ma, L., Hu, Y., Pal, S., ... & Hao, J. (2024). Enhancing logical reasoning in large language models through graph-based synthetic data. arXiv preprint arXiv:2409.12437.

**Publisher's Note** Springer Nature remains neutral with regard to jurisdictional claims in published maps and institutional affiliations.